\documentstyle[11pt,moriond,epsfig]{article}

\def\Journal#1#2#3#4{{#1} {\bf #2}, #3 (#4)}

\def\LL{{\ell_{_H}}^2}
\def\L{{\ell_{_H}}}
\def\Ns{N_\Phi}
\def\T{\mathcal{T}\mathit}
\def\m{{\tt m}}
\def\prb{{\em Phys. Lett.}  B}
\def\prl{\em Phys. Rev. Lett.}
\def\nature{\em Nature}
\def\be{\begin{equation}}
\def\ee{\end{equation}}
\def\bea{\begin{eqnarray}}
\def\eea{\end{eqnarray}}

\begin{document}

\title{Can Quasi-Particles Tunnel Through A Barrier?}

\author{Elad Shopen$^1$, Yuval Gefen$^2$, and Yigal Meir$^{1,3}$}

\address{$^1$  Physics Department, Ben-Gurion University, Beer Sheva 84105, Israel}
\address{$^2$ Department of Condensed Matter Theory, The Weizmann Institute of Science, Rehovot 76100, Israel}
\address{$^3$  The Ilse Katz Center for Meso- and Nano-scale Science and
Technology, Ben-Gurion University, Beer Sheva 84105, Israel}

\maketitle\abstracts{This is a qualitative description of a
systematic analysis carried out by us \cite{Shopen}. We address
the question of whether fractionally charged particles, in the
context of the fractional quantum Hall effect (FQHE) can tunnel
through a potential barrier (around which the density of the
quantum Hall liquid is practically zero). Setting the barrier in a
multiply-connected FQHE geometry removes the "global constraint"
which prohibits such tunnelling. We have performed a microscopic
analysis of adiabatic charge pumping and tunnelling in a torus
geometry. Below we summarize our analysis in semi-qualitative
terms. We also propose a setup-- different from the torus
geometry-- amenable to experimental verification.}

\section{Motivation}

A common approach for observing elementary charge carriers within
fractional quantum Hall effect (FQHE) systems is through
tunnelling.  It has been pointed out \cite{QPversEL,gefenless}
that QP tunnelling is distinctly different from electron
tunnelling. Perturbative renormalization-group analysis
\cite{kanefish} has indicated that in the weak backscattering
limit interedge tunnelling through the FQHE liquid is dominated by
QP tunnelling. These predictions have been confirmed by
experiments \cite{experiments}. In the opposite limit of strong
backscattering (nearly disconnected FQHE systems coupled by weak
tunnelling through an insulator), the same RG analysis would have
predicted that tunnelling should be dominated again by QP
tunnelling. Common wisdom, however, has it that in that limit only
electron tunnelling is possible. The rationale for that goes as
follows: consider two FQHE puddles weakly connected through
tunnelling. The total number of electrons on each puddle
($N_R,N_L$ respectively) is a (nearly) good quantum number; hence
it must be an integer. QP tunnelling would render this number
non-integer, therefore such a process must be
excluded\cite{footnote1}.

Our starting point here is to note that there are setups where the
above mentioned "global constraint" (i.e. the number of electrons
on each side of the barrier being an integer) does not exclude
a-priori QP tunnelling through a potential barrier. The common
wisdom alluded to above needs then to be re-examined. Studying
these setups is particulary interesting in view of recent
experimental results \cite{recent experiments} which suggest the
coexistence of both electron and QP tunnelling under rather strong
backscattering conditions. Throughout this paper we will refer to
the $\nu=1/m$ FQHE.

\section{Annular Geometry}

\begin{figure}
\begin{center}
\psfig{figure=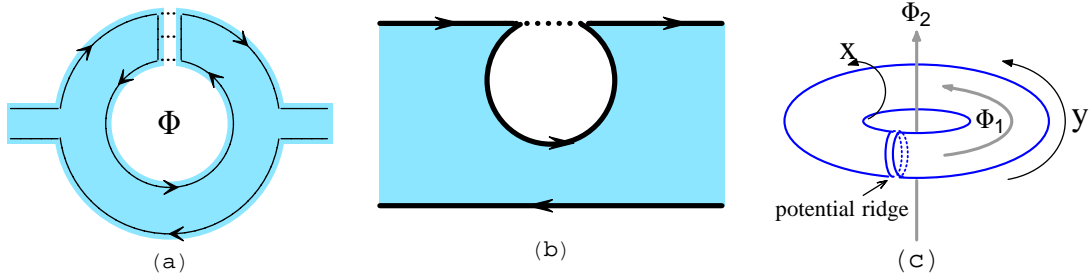,height=1.5in} \caption{Setups
for studying possible QP tunnelling through an insulator. (a) An
annular geometry. Solid lines indicate the directions of the edge
currents; dotted lines-- tunnelling trajectories. The shaded area
represents the $\nu=1/m$ FQHE liquid. (b) An equivalent geometry
in terms of edge and tunnelling currents. (c) A torus geometry.
$\Phi_1, \Phi_2$ are the two gauge fluxes. \label{fig:annulus}}
\end{center}
\end{figure}

The simplest geometry, probably most amenable to experimental
verification, is that of the annulus shown in
Fig.~\ref{fig:annulus}.a.

One of the arms of the annulus includes a tunnel barrier void of
Hall liquid. Tunnelling may then occur through this barrier. We
now present a few qualitative insights concerning this setup. The
obvious approach to study this system is by employing the chiral
Luttinger liquid model to describe the low energy dynamics at the
edges. Fig.~\ref{fig:annulus}.b shows chiral edges which are
topologically equivalent to the original system. Scattering due to
the tunnelling barrier (dashed line) is always forward. The zero
frequency transmission probability is therefore $T=1$ implying
that the (zero frequency) quantum shot-noise spectrum $S_2=V
(e^2/h)T(1-T)$ vanishes. Finite frequency noise may, nevertheless,
be generated. The mechanism to this is that the "loop" at the
barrier may be considered as a capacitor which gets stochastically
charged and discharged. In other words, fluctuations in the form
of charged wave-packets coming from left may follow several
windings around this loop before leaking to the right-hand-side of
the edge. While the d.c. conductance will {\it not} be sensitive
to a Aharonov-Bohm (AB) flux threading the loop (the zero
frequency transmission is anyway $1$), the finite frequency noise
{\it will} be sensitive to that flux. We expect the flux
periodicity of the noise to depend on whether the tunnelling is
dominated by quasi-particles or electrons.

\section{Charge and Aharonov-Bohm Periodicity}

The subtle relation between the charge of the elementary carriers
(or the elementary charge participating in tunnelling) and the AB
periodicity in multi-connected structured has been elucidated by
Thouless and Gefen \cite{gefenless} for the case of an  annular
geometry. In general the electronic ground state (in the FQHE
regime) may be multiply degenerate, forming a structure of
intersecting flux periodic minibands. Tunnelling gives rise to
avoided crossing, cf. Fig~\ref{fig:EnergyVersFlux}.a where the
$m=3$ case is depicted. This figure is quite generic, as shown by
our present analysis. Quasi-particle-tunnelling-generated minigaps
give rise to $\Phi_0 \equiv hc/e$ periodicity. If only electron
tunnelling is permitted the corresponding gap gives rise to $3
\Phi_0$ periodicity. Note that with $e^* = e/3$ this is equivalent
to $hc/e^*$ periodicity. Thus, the flux periodicity is a clear-cut
mean to identify the charge associated with tunnelling.

\section{Formulating the Challenge  and Working Out the Problem --  an
Outline of the Analysis}

\begin{figure}
\begin{center}
\psfig{figure=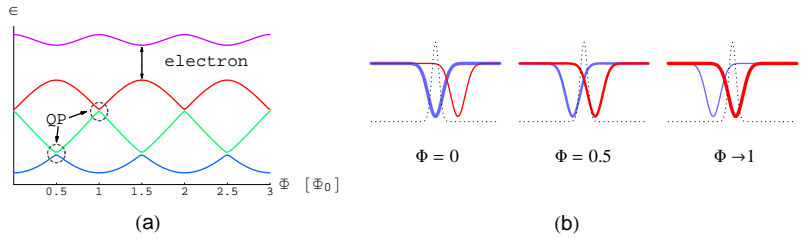,height=1.5in} \caption{Relation
between charge and Aharonov-Bohm periodicities. (a) Energies of
selected many-body configurations as function of AB flux. Finite
matrix elements for QP tunnelling give rise to gaps in the
adiabatic spectrum ("avoided crossings"), rendering the adiabatic
variation of the ground-state energy vs. flux $\Phi_0$-periodic.
In the absence of such matrix elements electron tunnelling gives
rise to a $m \Phi_0$-periodicity. (b) Density profiles (solid
lines) of two many-body configurations on the torus. These
configurations slide continuously as the flux $\Phi_1$ is modified
adiabatically. The minimum energy configuration shifts from  the
first to the second configuration, depending on whose density
minimum coincides with the  potential ridge (dotted line).
\label{fig:EnergyVersFlux}}
\end{center}
\end{figure}

The question at the heart of the dilemma here is whether
fractionally charged quasi-particles can indeed tunnel through a
potential barrier. Should the answer to this be in the
affirmative, the next step is to study the {\it magnitude} of the
quasi-particle tunnelling amplitude, $\T_{qp}$, as function of the
relevant parameters, and compare it with  the amplitude for
electron tunnelling, $\T_e$. In view of
Fig~\ref{fig:EnergyVersFlux}.a, these amplitudes are proportional
to the respective QP and electron gaps indicated there. As was
emphasized above the suppression of $\T_{qp}$ will lead to the
change of the flux periodicity from $\Phi_0$ to $m\Phi_0$. A
particularly intriguing question here is the role of impurities.
It turns out that in general their presence enhances the above
tunnelling amplitudes. But this will be discussed elsewhere.

We now present a qualitative  description of {\bf the main steps
of the analysis} carried out by us:
\\
(i) We recall the structure and analytic properties of the
lowest-Landau-level single electron wave-functions defined on the
surface of a torus (whose periodic coordinates are $x, y$) in the
presence of a strong perpendicular magnetic field ($N_{\Phi}$ flux
quanta). Likewise we review the many-body wave-functions in the
FQHE regime\cite{Yoshioka,Haldane1}. We note that these
wave-functions depend on the gauge fluxes, $\Phi_1$ and $\Phi_2$,
threading the torus (cf. Fig.\ref{fig:annulus}.c).
\\
(ii) We generalize the many-body wave-functions to include $N_h$
localized quasi-holes at points $z_{01}, z_{02}, \ldots,z_{0
N_h}$. These coordinates can be thought of as the locations of
point-like impurities. Alternatively these are the quasi-holes
generated by inserting extra $N_h$ flux quanta through the
surface, on top of the $N_{\Phi}$ fluxons \cite{Haldane2} (the
number of electrons $N=(N_{\Phi}-N_h)/m$).
\\
(iii) Many-body wave-functions with $N_h$ localized holes are now
expressed  in terms of  $N_h$ "extended holes". The latter refers
to wave-functions in which certain single electron states (certain
quasi  momentum components) are excluded. Consecutive extended
holes form a "dry swath" (a minimum or even a zero-density regime)
of circular symmetry. There are a number of degenerate many-body
states characterized by the location of the dry swath.
\\
(iv) We now introduce, on top of the torus surface, a potential
ridge, Fig.~\ref{fig:annulus}.c. This ridge removes the degeneracy
of the many-body states. The wave-function with the dry swath
coinciding with the ridge has the lowest energy
(Fig~\ref{fig:EnergyVersFlux}.b, $\Phi=0$).
\\
(v) By adiabatically varying the gauge flux $\Phi_1$ we cause the
incompressible liquid (and the dry swath) represented by the
many-body wave-function $\Psi_n$ to rigidly slide around the torus
in the $y$-direction. Note that this sliding implies that the
liquid (the incompressible "sea") may climb up and down the ridge.
This is still not what we would call tunnelling.  The trace of the
energy vs. $\Phi_1$ ($E_n(\Phi_1)$) is modulated with the flux.
Also -- as the swath of $\Psi_n$ moves away from the ridge, the
swath of another state, $\Psi_{n+1}$, replaces it, resulting in
$E_n(\Phi_1)$ and $E_{n+1}(\Phi_1)$ intersecting. The degeneracy
of these two states at the intersection point is not removed:
$\Psi_n$ and $\Psi_{n+1}$ differ in their total quasi momentum
(TQM), and the circularly symmetric ridge cannot provide a matrix
element to remove the degeneracy between them.
\\
(vi) We now break the local circular symmetry of the problem by
introducing an extra potential (e.g., an additional
$\delta$-function potential on top of the ridge). This may now
generate matrix elements among the sliding states whose energy
traces intersect. These matrix elements, leading to gaps in the
spectrum (cf. Fig~\ref{fig:EnergyVersFlux}.a) are manifestations
of tunnelling.
\\
(vii) Physically the tunnelling can be understood in the following
terms. By adiabatically varying $\Phi_1$ the rigid electron
configuration associated with $\Psi_n$ is pushed (pumped) around
the torus. Increasing $\Phi_1$ by $\Phi_0$ will push the state by
one guiding center, {\it i.e.} will effectively push a QP of
charge $1/m$ across the swath. This will result in another
many-body state, $\Psi_{n+1}$. If the circular-symmetry breaking
potential alluded to above generates a (significant) finite matrix
element between these two states, we identify the process as an
effective QP tunnelling across the ridge (corresponding to the QP
gap in Fig~\ref{fig:EnergyVersFlux}.a). Otherwise we need to
increase $\Phi_1$ by $m\Phi_0$, pushing the charge of an electron
across the ridge, to generate a (significant) matrix element
between $\Psi_n$ and $\Psi_{n+m}$ (an electron gap, cf.
Fig~\ref{fig:EnergyVersFlux}.a).
\\
(viii) Once the magnitudes of the electron  and the QP tunnelling
matrix elements are evaluated, we can find the crossover (as
function of the torus geometrical characteristics and the
thickness of the ridge potential barrier) between the two
processes.

\section{Results}

\begin{figure}[ht]
\begin{center}
\psfig{figure=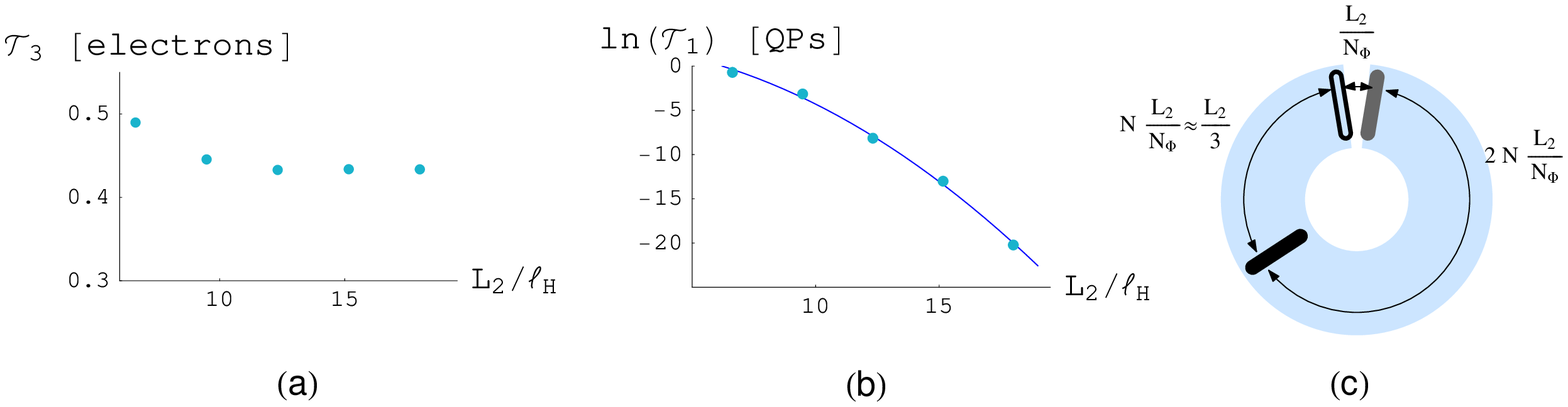,height=1.7in}
\caption{Dependence of tunnelling amplitudes on system's size.
Data shown refer to a single extended hole, $N=2,\ldots,6$.  (a)
For electrons $\T_3$ is practically constant while (b) for QPs
$\T_1$ strongly decreases $\sim e^{-\alpha L_2^2}$ with the linear
size of the torus (all lengths are measured in units of $\L$, the
magnetic length). (c) Insight on the difference between electron
and QP tunnelling: the empty box (just to the left of the
potential ridge) denotes the location of the extended hole of the
initial state. The tunnelling of a QP (an electron) involves
initial and final many-body states, whose TQMs  differ by  $N$
($3N$). The black (grey)  box denotes the location of the extended
hole following QP (electron) tunnelling. The respective matrix
elements involve the overlap of two single particle states
(Gaussians) whose distance is $N L_2/\Ns \sim L_2/3$ (for
electrons $3 N L_2 /\Ns ~\mbox{mod} (L_2) = L_2 N_h/\Ns \ll L_2$).
Thus for QPs the tunnelling matrix element scales as
$e^{-(L_2/3)^2/4}$, while for electrons $\sim e^{-(L_2/\Ns)^2/4}$
independent of $L_2$ (note that $L_2 \propto \Ns$).
\label{fig:tunnelling}}
\end{center}
\end{figure}

Below we present some of our results. First we observe (in our
multiply-connected geometry, no "global constraint") a
non-vanishing amplitude for QP tunnelling. The interesting
question then is whether the effect is mesoscopic or is it also
observable in the thermodynamic limit. This can be answered by
studying how the tunnelling amplitudes for electrons and QPs
depend on system's size.  The result is quite striking: we have
found that tunnelling of electrons is practically unaffected by
increasing the size of the system (keeping the potential barrier
unchanged), while the tunnelling amplitude for QPs vanishes
(Gaussian-like). Hence tunnelling of QPs \emph{is} a mesoscopic
effect. Such tunnelling involves the entire electron sea and  is a
manifestation of the overlap between the initial  and   final
states of this sea. Let us show this more quantitatively: the
tunnelling amplitude is given by $\T_k \equiv
<\Psi_n|V|\Psi_{n+k}>$, where $\Psi_n$ is a Laughlin correlated
wave function with extended holes (the dry swath whose center is
the $n$-th single particle state); $V$ is the symmetry breaking
potential, taken for convenience to be a Dirac delta-function
$\delta(x)$ (this specific choice does not modify our qualitative
results); $k=1, m$ correspond to QP and electron tunnelling
respectively. The size of the system may be increased by
increasing the number of particles and flux quanta, keeping the
occupation at the sea $1/m$. We first present results for a narrow
circular ridge ($N_h=1$). Fig~\ref{fig:tunnelling}.a-b depict our
results for $N=2,\ldots,6$. We find for $m=3$
\begin{eqnarray}%\label{}
    \mbox{(QP)}&\T_1& \sim e^{-\alpha L_2^2/ \LL}  \\
    \mbox{(electron)}&\T_3& \sim \mbox{constant (as function of
    $L_2$)}
    ~,
\end{eqnarray}
with $\alpha \approx 1/14$. Here $L_2$ is the torus length and
$\L=\sqrt{\hbar c/e B}$ the magnetic length.

It is clear that tunnelling of electrons is system size
independent (i.e., it does not depend on the length -- in the
$y$-direction --  of the sea), as opposed to QP tunnelling. While
QP tunnelling through the barrier does exist, it is suppressed
with the linear size of the electron sea  \emph{as if} the QP
tunnelling takes place through the sea.

How can one understand this result qualitatively? Let us begin
with QPs. The extension to electrons is then straightforward. The
tunnelling amplitude $\T_1$ involves $\Psi_n$ and $\Psi_{n+1}$.
The difference in TQM between these two states is $N$ (the latter
being the number of electrons): $\Psi_{n+1}$  is obtained from
$\Psi_n$ by increasing the quasi  momentum of each single particle
component of the many-body wave-function by one, rendering the
overall change of the TQM  equal to the number of electron, $N$.
Since the potential $V$ is a single particle operator,
$<\Psi_n|V|\Psi_{n+1}>$ is proportional to the product of  single
particle states whose quasi momentum difference is $N$. Let us
approximate a single particle state by a Gaussian $\sim
e^{-(y-y_{c})^2/2 \LL}$, with the guiding center at $y_{c}$.
Adjacent guiding centers are distanced $L_2/\Ns$ from one another.
The guiding centers alluded to above are a distance $N (L_2/\Ns)
\approx L_2/\m$ apart ($\m=3$ is the inverse of the filling
factor, $\Ns=\m N+1$). Subsequently $<\Psi_n|V|\Psi_{n+1}> \sim
\int e^{-y^2/2 \LL} e^{-(y-L_2/\m)^2/2 \LL}dy \sim
e^{-(L_2/2\m)^2}$. $\T_1$ was previously calculated by Auerbach
\cite{assa} for a cylindrical geometry vis-a-vis tunnelling of QPs
through a quantum Hall liquid. The present discussion applies for
Auerbach setup as well. This estimate verifies the Gaussian-like
decrease, however the actual decay is stronger: $\m$ rather than
$\m^2$ in the exponential. This is due to the normalization and
combinatorial factors of the many-body wave-functions  involved.
It can be \emph{pictorially} described as a QP performing $\m$
hops each a distance  $L_2/\m$. As each hop involves a decaying
factor of $e^{-(L_2/2\m)^2}$, one reproduces the
$e^{-(L_2/2)^2/\m}$ factor of Auerbach. We speculate that the
$1/14$ factor alluded to above approaches $\alpha=1/12$ at larger
values of $L_2$.

Electrons tunnelling involves many-body states $\Psi_n$ and
$\Psi_{n+m}$, with TQM difference of $mN$. The guiding centers of
the single particle wave-functions whose overlap we address are a
distance $(mN)(L_2/\Ns)$ apart. But as we are studying a torus,
distances are defined modulo $L_2$, hence $(mN)(L_2/\Ns)
modulo(L_2) =L_2/\Ns$, which is system size independent (recall
that $\Ns=mN+1=L_1 L_2/\LL$). The physics discussed here is
depicted (for $m=3$) in Fig~\ref{fig:tunnelling}.c.

Extending our study to thicker ridges (i.e., increasing the number
of extended holes) supports the above conclusions- tunnelling of
QPs practically does not depend on $L_{barrier}$, the barrier
thickness, while electron tunnelling decreases as
$e^{-(L_{barrier}/2)^2}$. A more careful analysis should account
for the fact that electrons can tunnel through both- the barrier
and the quantum Hall liquid, the latter depends Gaussian-like on
the liquid length $L_{liquid}$, $\sim e^{-(L_{liquid}/2)^2}$. This
latter correction to electron tunnelling is sub-dominant to QPs
tunnelling \cite{assa}.

\begin{figure}[ht]
\begin{center}
\psfig{figure=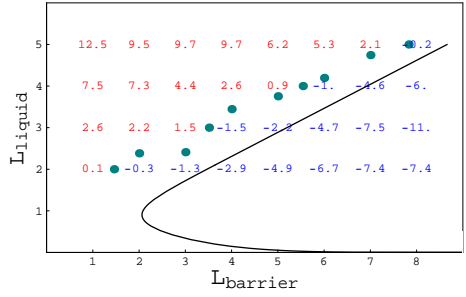,height=1.5in} \caption{$\chi
\equiv ln(\T_3/\T_1)$ (ratio of electron to QP tunnelling
amplitudes) for the torus. Numerically computed values of $\chi$
are indicated. The lengths of the barrier and the "sea" are
measured in units of $L_2/\Ns$. Thick dots mark the (numerically
interpolated) positions of $\chi=0$. The solid curve marks the
crossover between electron-to-QP-dominated tunnelling and is in
agreement with our numerical data. This crossover curve is based
on a simplified picture whereby the electron tunnel amplitude is
the sum of tunnelling processes through the barrier and the sea
which scale respectively as $\sim e^{-L_{barrier}^2/4}$ and $\sim
e^{-L_{liquid}^2/4}$; QP tunnelling scale as $\sim
e^{-L_{liquid}^2/4 \m}$ ($\m=1/\nu$). \label{fig:crossover}}
\end{center}
\end{figure}

A rough estimate for the crossover between electron and QP
dominated tunnelling regimes, writing $\T_1 =
e^{-(L_{liquid}/2)^2/\m}$ and $\T_3
=e^{-(L_{liquid}/2)^2}+e^{-(L_{barrier}/2)^2}$ (this ignores
pre-exponential prefactors). Fig~\ref{fig:crossover} depicts the
crossover curve, in fair agreement with a numerical calculation of
$ln(\T_3/\T_1)$. We believe that the experimental study of such an
electron-to-QP crossover in multiply connected systems (e.g., an
annulus) is now feasible.

Upon completion of this work we have learned of the manuscript by
M. Helias and  D. Pfannkuche,  cond-mat/0403126. In this work
numerical evidence for the occurrence of quasi-particle tunnelling
near a potential saddle-point is reported.

\section{Acknowledgment}

We acknowledge useful discussions with F.D.M. Haldane, B.I.
Halperin, M. Heiblum, A.D. Mirlin and D.J. Thouless.  This work
was supported in part by the US-Israel binational science
foundation (BSF), by the Israel Science Foundation (ISF) of the
Israel Academy of Science, and by the Alexander von Humboldt
Foundation (through the Max Planck Award). Y.M. was supported at
the WIS by the Albert Einstein Minerva Center (BMBF).


\begin{thebibliography}{99}
\bibitem{footnote1} To be more precise, one may tune the relative
chemical potentials of the two puddles such that the last occupied
discrete level on, say, the right-hand-side, is aligned with the
first vacant level on the left; indeed, $N_L, N_R$ need not be
integers then. Nevertheless there is no indication that as
function of the relative gate voltage they show a stepwise
structure at $1/m, 2/m, \Pi...$ , which would be a clear
indication of QP tunnelling.
\bibitem{Haldane2} Many-body states on a torus with a
single localized hole have been introduced in Ref.
\cite{Haldane1}.
\end{thebibliography}
\end{document}